\documentclass[prl,twocolumn,showpacs,floatfix,superscriptaddress]{revtex4}

\usepackage{amsmath,amssymb,amsthm}
\usepackage{times}
\usepackage{graphicx}
\usepackage{bm}
\usepackage[dvips]{color}

\begin{document}

\title{Weak and strong typicality in quantum systems}

\author{Lea F. Santos}
%\email{lsantos2@yu.edu}
\affiliation{Department of Physics, Yeshiva University, New York, NY 10016, USA}
\author{Anatoli Polkovnikov}
%\email{asp@bu.edu}
\affiliation{Department of Physics, Boston University, Boston, MA 02215, USA}
\author{Marcos Rigol}
%\email{mrigol@physics.georgetown.edu}
\affiliation{Department of Physics, Georgetown University, Washington, DC 20057, USA}
\affiliation{Physics Department, The Pennsylvania State University, 104 Davey Laboratory, 
University Park, Pennsylvania 16802, USA}

\pacs{05.70.Ln, 05.30.-d, 05.45.Mt, 02.30.Ik}

\begin{abstract}
We study the properties of mixed states obtained from eigenstates of many-body lattice Hamiltonians after 
tracing out part of the lattice. Two scenarios emerge for generic systems: (i) the diagonal entropy becomes 
equivalent to the thermodynamic entropy when a few sites are traced out (weak typicality); 
and (ii) the von Neumann (entanglement) entropy becomes equivalent to the thermodynamic entropy  
when a large fraction of the lattice is traced out (strong typicality). Remarkably, the results for 
few-body observables obtained with the reduced, diagonal, and canonical density matrices are very similar
to each other, no matter which fraction of the lattice is traced out. Hence, for all physical quantities 
studied here, the results in the diagonal ensemble match the thermal predictions.
\end{abstract}

\maketitle

Despite advances, the emergence of thermodynamics from quantum mechanics is still a subject under debate.  
How to derive, from first principles, proper ensembles leading to the basic thermodynamic relations is not 
yet entirely clear. According to the concept of canonical typicality~\cite{typicality}, the reduced density 
matrix of a subsystem of most pure states of many-particle systems is canonical. The proof of this statement 
requires a partition of the original system into a small ``system'' and a large ``environment''. However, suppose, 
for example, that our universe is in a pure state and that we trace out only a finite number of degrees 
of freedom. Can we describe the rest of the universe using statistical mechanics? How much one needs to trace out, 
how well the notion of canonical typicality works in finite systems, and which quantities will be more or less 
affected by the fraction of the original system traced out are questions that have received little attention. 
They are the more pressing given the progress made in experiments with ultracold gases~\cite{OpticalLattices}.

We can discuss these issues in the context of entropy. If one takes an eigenstate of a generic many-body Hamiltonian 
and traces  out the ``environment'' (${\cal E}$), the grand canonical (GC) ensemble is the appropriate ensemble to 
describe the system (${\cal S}$) that is left, where the total energy and number of particles fluctuate. One immediately 
realizes that the GC-entropy, $S_\text{GC}$, must be different from the von Neumann entropy, $S_\text{vN}$, if only 
a small fraction of the original system is traced out. In this case, $S_\text{vN}$ is extensive in the size of the 
environment, instead of in the size of the system as $S_\text{GC}$. This follows from the fact that, for a 
composite system ${\cal S+E}$ in a pure state $\hat{\rho}=|\Psi\rangle\langle \Psi|$, the reduced von Neumann 
entropy is defined as
\begin{equation}
S_\text{vN}\equiv -\mbox{Tr}_{\cal S}\left[ \hat{\rho}_{\cal S} \ln\hat{\rho}_{\cal S} \right]
           \equiv -\mbox{Tr}_{\cal E}\left[ \hat{\rho}_{\cal E} \ln\hat{\rho}_{\cal E} \right],
\label{SvN}
\end{equation}
where the Boltzmann constant is set to unity and the reduced density matrix $\hat{\rho}_{\cal S}=\mbox{Tr}_{\cal E} \hat{\rho}$ 
and similarly $\hat{\rho}_{\cal E}=\mbox{Tr}_{\cal S} \hat{\rho}$. $S_\text{vN}$ has been widely used to measure the entanglement 
in bipartite systems~\cite{Mejia2005}. If the pure state is separable, $\hat\rho=\hat{\rho}_{\cal S}\otimes\hat{\rho}_{\cal E}$, 
then $S_\text{vN}=0$, while maximum entanglement leads to $S_\text{vN}=\ln \cal{D}$, where $\cal{D}$ is the smallest 
dimension of the two subsystems. The source of the disparity between $S_\text{vN}$ and $S_\text{GC}$ is the information 
present in the off-diagonal elements of the reduced density matrix, which is not contained in the thermodynamic ensemble. 
When a large fraction of the original system is traced out, the equivalence between $S_\text{vN}$ and $S_\text{GC}$ is expected 
(canonical typicality). However, up to our knowledge, it has not been demonstrated for realistic systems.

Now suppose that instead of tracing out the environment, we physically cut it off and let the remaining system relax to 
equilibrium. We can then ask how much one needs to cut for the entropy of the reduced system to become equivalent to $S_\text{GC}$
(if ever). After relaxation, this system is described by the diagonal ensemble~\cite{rigol08,rigol09STATa} obtained by writing 
the reduced density matrix ($\hat{\rho}_{\cal S}$) in the energy eigenbasis. The proper thermodynamic entropy in this case
has been argued to be the diagonal entropy \cite{Polkovnikov2011,Santos2011PRL}, defined as
\begin{equation}
S_d \equiv -\sum_{n} \rho_{nn} \ln (\rho_{nn}),
\label{Sd}
\end{equation}
where $\rho_{nn}$ are the diagonal elements of the density matrix. $S_d$ counts logarithmically the number of 
energy eigenstates which are occupied in ${\cal S}$. Unlike $S_\text{vN}$, $S_d$ is extensive in the 
system size even if we trace a very small environment, e.g., a single degree of freedom. Cutting off the environment 
is equivalent to a sudden quench; it introduces to the system the (nonextensive) energy uncertainty $\delta E_{\cal S}$. 
This uncertainty implies that if all the eigenstates of the subsystem Hamiltonian $\hat{H}_{\cal S}$ within the window 
$\pm\delta E_{\cal S}$ are occupied with roughly the same weights \cite{weights}, then 
$S_d\approx \log \Omega(E_{\cal S})\delta E_{\cal S}$, where $\Omega(E_{\cal S})$ is the density of states in 
${\cal S}$. Up to subextensive corrections, $S_d$ would then coincide with the thermodynamic entropy of ${\cal S}$.

In this Letter, we study the properties of mixed states obtained from eigenstates of a many-body lattice Hamiltonian 
after tracing out or cutting off an increasingly large environment. One of our goals is to understand the structure 
of the remaining reduced density matrix and if it ever becomes thermal, which we monitor using $S_\text{vN}$, $S_\text{GC}$, 
and $S_d$. We find that, for nonintegrable systems, $S_d$ approaches $S_\text{GC}$ after cutting off a few (possibly one) sites, 
i.e., our hypothesis above is verified. $S_\text{vN}$, on the other hand, 
remains different from $S_d$ and $S_\text{GC}$ until a large fraction of the lattice is traced out \cite{josh}. This 
motivates us to distinguish between the conventional (or strong) typicality $S_\text{vN} \cong S_d\cong S_\text{GC}$ and 
a weaker typicality in the sense that $S_\text{vN}\neq S_d\cong S_\text{GC}$. The latter implies that {\it only} the diagonal 
part of the density matrix of the reduced system in the energy eigenbasis exhibits a thermal structure. 

Our results then show that the diagonal entropy satisfies the key thermodynamic relation:
\begin{equation}
{\partial S_{\cal S}\over \partial E_{\cal S}}={\partial S_{\cal E}\over\partial E_{\cal E}}={1\over T},
\end{equation}
where $E_{\cal S}$ and $E_{\cal E}$ are the energies of the subsystems. This follows from the fact that $S_d$ coincides 
with the thermodynamic entropy for ${\cal S}$ and ${\cal E}$ simultaneously. In contrast, $S_\text{vN}$ cannot satisfy this 
equality, as one can see by considering ${\cal E\gg S}$. In this case, $S_\text{vN}=S_{\cal S}=S_{\cal E}$ 
is proportional to the size of ${\cal S}$, while $E_{\cal E}$ is proportional to the size of ${\cal E}$, so 
$\partial S_{\cal E}/\partial E_{\cal E}\to 0$. 

Another of our main goals in this work is to understand the description of few-body observables in the mixed states 
obtained by the two procedures mentioned before. For that, we study their expectation values as given by the reduced 
density matrix, the diagonal ensemble, and the GC ensemble. The results for the first two are similar even if  
very few sites are traced out and the system sizes are small. This suggests that either tracing out part of the original 
system or removing the same number of sites and waiting for the reduced system to relax leads to the same results, up to 
non-extensive boundary terms. For all practical purposes both procedures are then equivalent. The agreement with the GC 
expectation values is also good and improves with increasing system size. This implies that, for few-body observables, 
only weak typicality is needed to observe thermal behavior in experiments.

%%%%%%%%%%%%%%%%%%%%%%%%%%%%%%%%%%%%%%%%%%%%%%%%%%%%%%%%%%%%%%%%%%%%%%%%%%%%%%%%%

{\em System.--} We study hard-core bosons in a one-dimensional lattice with open-boundary conditions described by
{\setlength\arraycolsep{0.5pt}
\begin{eqnarray}
&& \hat{H} =\epsilon \left(\hat{n}_1 -\frac{1}{2} \right)  \label{bosonHam} \\
&& +\sum_{i=1}^{L-1} \left[ -t \left(\hat{b}_i^{\dagger} \hat{b}_{i+1} + \textrm{H.c.} \right) +
V \left(\hat{n}_i -\frac{1}{2} \right)\! \left(\hat{n}_{i+1} -\frac{1}{2}\right) \right] 
\nonumber \\
&& +\sum_{i=1}^{L-2} \left[
-t' \left( \hat{b}_i^{\dagger} \hat{b}_{i+2} + \textrm{H.c.}\right)+
V' \left(\hat{n}_{i} -\frac{1}{2}\right) \! \left(\hat{n}_{i+2} -\frac{1}{2}\right)
\right], \nonumber
\end{eqnarray}
}where, $t$ and $t'$ [$V$ and $V'$] are nearest-neighbor (NN) and next-nearest-neighbor (NNN) hopping [interaction], 
$L$ is the chain size, and standard notation has been used~\cite{Santos2010PRE}. Symmetries, and therefore degeneracies, 
are avoided by considering 1/3-filling and by placing an impurity ($\epsilon\neq0$) on the first site. In what follows, 
$t=V=1$ sets the energy scale, $\epsilon=1/5$, and $t'=V'$. When $t'=V'=0$, the model is integrable~\cite{integrableAlcaraz}. 
As the ratio between NNN and NN couplings increases, the system transitions to the chaotic domain \cite{Santos2010PRE}. 
The results below depend on the regime (integrable vs nonintegrable), but not on specific values of the parameters.

For our calculations, we select an eigenstate $|\Psi_j\rangle$ of $\hat{H}$ (\ref{bosonHam}), with energy $E_j$ closest 
to $E = \sum_j E_j e^{-E_j/T}/\sum_{j} e^{-E_{j}/T}$ corresponding to an effective temperature $T$. 
Since $S_\text{vN}=S_d=0$, this can be seen as the most distant choice for a state with a thermodynamic entropy.
We then trace out a certain number of sites $R \leq 2L/3$, on the right side of the chain, and study 
the entropies and observables of the reduced system. The reduced density matrix $\hat{\rho}_{\cal S}$ describing the remaining 
system consists of different subspaces each with a number $N \in [{\rm max}(L/3-R,0),L/3]$ of particles.

%%%%%%%%%%%%%%%%%%%% ENTROPY RESULTS %%%%%%%%%%%%%%%%%%%%%%%%%%%%%%%%%%%
{\em Entropies.--} The von Neumann, diagonal, and GC entropies are given by Eq.~(\ref{SvN}), Eq.~(\ref{Sd}), and
\begin{equation}
S_\text{GC} = \ln \Xi + \frac{E_{\cal S}-\mu N_{\cal S}}{T_\text{GC}},
\label{Sgc}
\end{equation}
respectively. In Eq.~\eqref{Sgc}, $\Xi=\sum_n e^{(\mu N_n - E_n)/T_\text{GC}}$ is the grand partition function, 
$T_\text{GC}$ is the GC temperature, $\mu$ is the chemical potential, and 
$E_{\cal S}= \mbox{Tr} [\hat{H}_{\cal S} \hat{\rho}_{\cal S}]$ and 
$N_{\cal S}= \mbox{Tr} [\hat{N}_{\cal S} \hat{\rho}_{\cal S}]$ are, respectively, the average energy and number 
of particles in the remaining subsystem ${\cal S}$.

\begin{figure}[!t]
\includegraphics[width=0.47\textwidth]{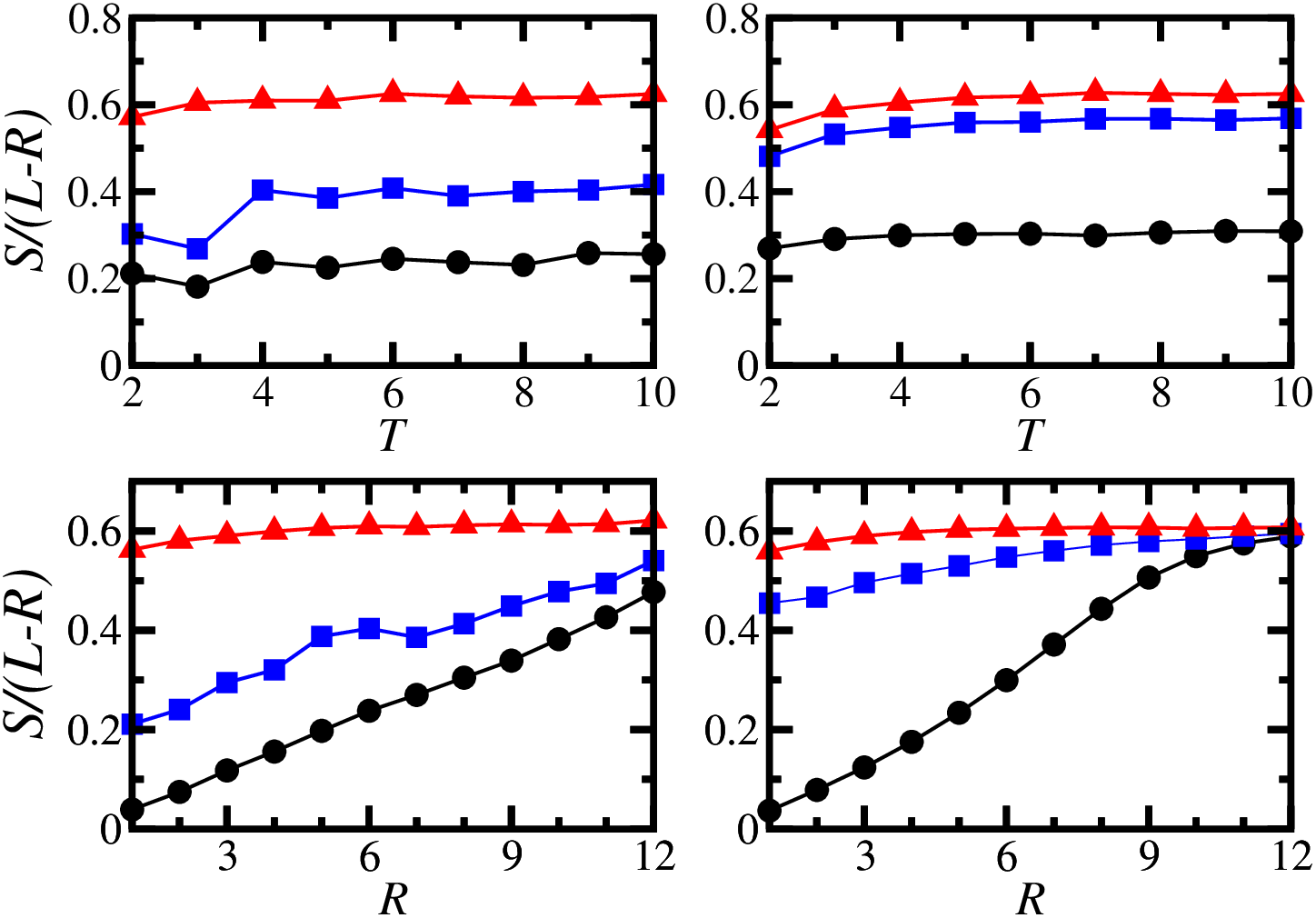}
\vspace{-0.25cm}
\caption{(Color online) (a),(b) Entropies per site vs temperature for $R=L/3$. 
(c),(d) Entropies per site vs $R$ for a fixed temperature; $T=4$. 
(a),(c) $t'=V'=0$ (integrable); (b),(d) $t'=V'=0.32$ (chaotic). All panels: $L=18$.}
\label{fig:entropyCut}
\end{figure}

{\em Results.--} In Fig.~\ref{fig:entropyCut}, we show results for $S_\text{vN},\ S_d$, and $S_\text{GC}$ in 
the integrable [(a),(c)] and chaotic [(b),(d)] domains. Larger fluctuations are seen in Figs.~\ref{fig:entropyCut}(a) 
and \ref{fig:entropyCut}(c) as characteristic of the integrable regime. Figs.~\ref{fig:entropyCut}(a) and 
\ref{fig:entropyCut}(b) show the entropies vs different eigenstates, which are 
increasingly away from the ground state ($T$ increases), when 1/3 of the sites are 
traced out. In the chaotic regime, and for all states selected, one can see that $S_d$ is close 
to $S_\text{GC}$, while $S_\text{vN}$ is quite far. This hints a thermal structure in the 
diagonal part of the reduced density matrix in the energy eigenbasis. 

In Figs.~\ref{fig:entropyCut}(c) and \ref{fig:entropyCut}(d) we show results for a fixed $T$ as an increasingly 
larger fraction of the original system is traced out. One can see again that in the chaotic limit $S_d$ is much 
closer to $S_\text{GC}$ than to $S_\text{vN}$, even when very few sites are traced out. However, in both regimes, 
all entropies approach each other as the fraction of sites traced out increases. 
For the lattice sizes considered here, we need to trace out more than one half of the chain for the effects of 
the off-diagonal elements of the density matrix to become irrelevant in $S_\text{vN}$, leading this entropy to 
finally approach $S_d$ and $S_\text{GC}$.

\begin{figure}[!t]
\includegraphics[width=0.47\textwidth]{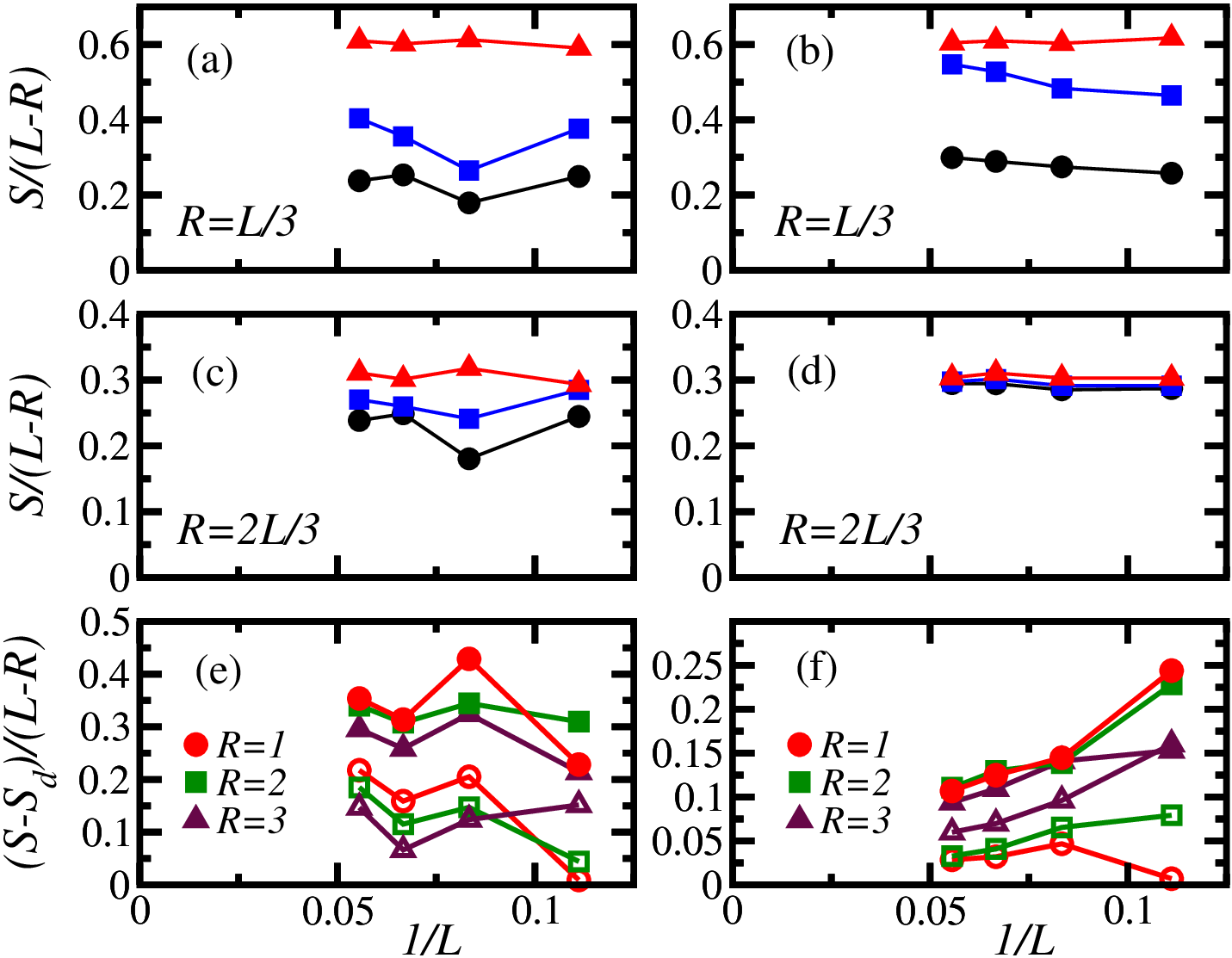}
\vspace{-0.25cm}
\caption{(Color online) (a)--(d) Entropies per site vs $1/L$ for $T=4$ and a fixed ratio $R/L$ (indicated). 
(e),(f) Full (empty) symbols: $S_\text{GC} - S_d$ ($S_\text{mc} - S_d$) per site vs $1/L$ for $R=1$, 2, and 3. 
(a),(c),(e) $t'=V'=0$ (integrable); (b),(d),(f) $t'=V'=0.32$ (chaotic).}
\label{fig:entropyL}
\end{figure}

The results presented in Fig.~\ref{fig:entropyCut} were obtained for $L=18$, the largest system size that we can study
with full exact diagonalization. Figure~\ref{fig:entropyL} depicts the scaling of the entropies in both domains, 
integrable [(a),(c)] and nonintegrable [(b),(d)], and for $T=4$. In Figs.~\ref{fig:entropyL}(a) and 
\ref{fig:entropyL}(b), when $L/3$ sites are cut off, $S_d$ and $S_\text{GC}$ approach each other as $L$ increases, 
up to a possible non-extensive correction. This trend is seen for all systems we have studied in the chaotic regime 
\cite{supp}, and opens up a new question: could cutting off an infinitesimal part of the original system lead 
$S_d$ and $S_\text{GC}$ to be equal in the thermodynamic limit? In Figs.~\ref{fig:entropyL}(e) and \ref{fig:entropyL}(f), 
we show the difference between $S_\text{GC}$ and $S_d$ per site vs system size when tracing out one, two, or three sites. 
In the chaotic regime [Fig.~\ref{fig:entropyL}(f)], the results are consistent with a vanishing difference in the 
thermodynamic limit (even when cutting one site \cite{supp}). This finding is  reinforced with the 
empty symbols, which show the difference between the microcanonical entropy ($S_\text{mc}$) and $S_d$. In this case, 
finite size effects are significantly reduced, leading to a much better agreement between the two entropies, which 
further improves with $L$ \cite{micro}. Hence, one could argue that single eigenstates of 
generic many-body Hamiltonians have a thermodynamic entropy \cite{josh}. In Figs.~\ref{fig:entropyL}(a) and \ref{fig:entropyL}(b), one can also see that for $R=L/3$, the von Neumann entropy (per site) 
saturates to a different value from $S_d$ and $S_\text{GC}$ (per site), as the lattice size increases. As shown in 
Figs.~\ref{fig:entropyL}(c) and \ref{fig:entropyL}(d), it is only when $R>L/2$ that the three entropies become comparable. 
However, for our system sizes, 
this happens only in the chaotic regime.

Close to the integrable point 
[Figs.~\ref{fig:entropyL}(a), \ref{fig:entropyL}(c), and \ref{fig:entropyL}(e)], large fluctuations are observed 
for different values of $T$ and $t',V'$, which makes it difficult to draw general conclusions. 
Fluctuations are indeed expected to be 
larger in the integrable regime than in the chaotic one. In the chaotic regime (and away from the edges of the spectrum) 
all eigenstates of the Hamiltonian that are close in energy have (i) a similar structure, as reflected by the inverse 
participation ratio and information 
entropy in different bases \cite{Santos2010PRE}, and (ii) thermal expectation values of few-body observables 
\cite{eth,rigol08,rigol09STATa,neuenhahn_marquardt_10}. However, this is not the case close to integrability 
where most quantities fluctuate wildly between eigenstates close in energy
\cite{Santos2010PRE,rigol08,rigol09STATa,neuenhahn_marquardt_10}, and this is affecting our results here.

%%%%%%%%%%%%%%%%%%%% OBSERVABLE RESULTS %%%%%%%%%%%%%%%%%%%%%%%%%%%%%%%%
{\em Observables.--} An important question we are left to address is whether the extra information carried by the off-diagonal 
elements of the reduced density matrix is of relevance to quantities measured experimentally. We focus our analysis on few-body 
observables. Their expectation values from the reduced, diagonal, and grand canonical density matrices are given by
\begin{eqnarray}
&& O_\text{vN}= \mbox{Tr}[\hat{O} \hat{\rho}_{\cal S}],\hspace{0.4 cm} O_d=\sum_n \rho_{nn} O_{nn}, \label{Od} \\
&& O_\text{GC} = \frac{1}{\Xi} \sum_n O_{nn} e^{(\mu N_n - E_n)/T}  \label{Ogc},
\end{eqnarray}
respectively. Here, $O_{nn}=\langle \psi_{n}|\hat{O}|\psi_{n}\rangle$ and $|\psi_{n}\rangle$'s are the 
eigenstates of the Hamiltonian in the reduced system.

\begin{figure}[!t]
\includegraphics[width=0.47\textwidth]{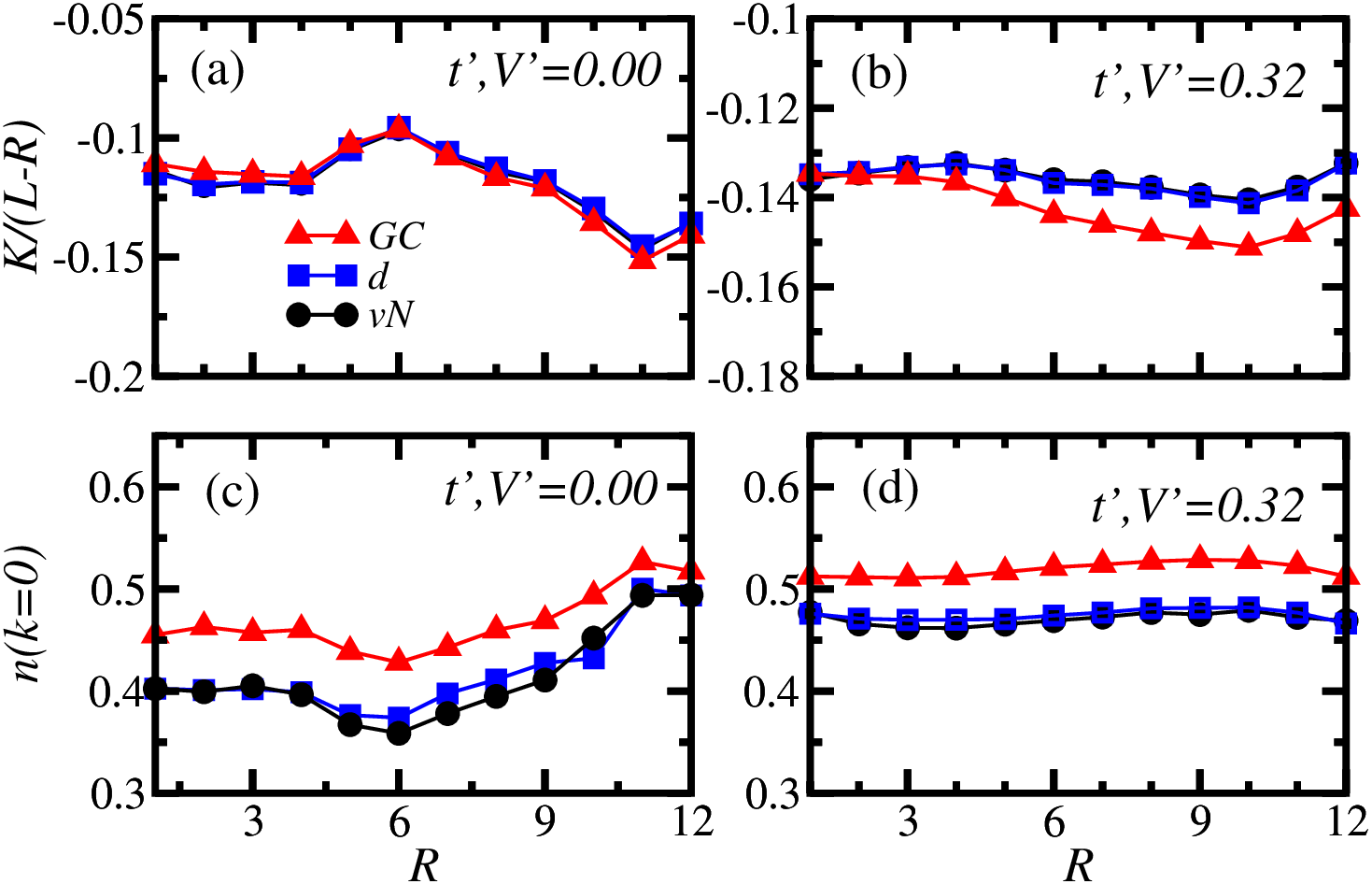}
%\vspace{-0.25cm}
\caption{(Color online) (a),(b) Kinetic energy per site and (c),(d) $n(k=0)$ vs $R$; $L=18$; $T=4$. 
(a),(c) $t'=V'=0$ (integrable); (b),(d) $t'=V'=0.32$ (chaotic).}
\label{fig:Ocut}
\end{figure}

{\em Results.--} Figure~\ref{fig:Ocut} shows results for the kinetic energy
\begin{equation}
\hat{K}=-t\sum_{i=1}^{L-R-1} \left( \hat{b}^\dagger_i \hat{b}_{i+1} + \textrm{H.c.} \right)
-t'\sum_{i=1}^{L-R-2} \left( \hat{b}^\dagger_i \hat{b}_{i+2} + \textrm{H.c.} \right),
\label{KE}
\end{equation}
and the momentum distribution function,
\begin{equation}
\hat{n}(k)=\frac{1}{L-R}\sum_{j,l=1}^{L-R} e^{i \frac{2\pi k}{L-R}(j-l)} \hat{b}^\dagger_j \hat{b}_{l},
\label{dist_n}
\end{equation} 
for an increasingly large fraction of sites traced out. The results obtained with the three density matrices are comparable, 
independently of the regime, the number of sites traced out, and $T$ (see~\cite{supp}). In particular, 
the fact that $O_\text{vN}$ and $O_d$ are so close demonstrates that both procedures, tracing sites out or cutting off part 
of the original system and then measuring the observables after relaxation, lead to similar outcome. The information contained 
in the off-diagonal elements of the reduced density matrix, which generates discrepancies between $S_\text{vN}$ and 
$S_d\cong S_{\rm CG}$, is therefore irrelevant to few-body observables. 
Our findings also open the interesting question: which experimentally measurable quantities (if any) could distinguish between 
tracing sites out or cutting them off?

We note that, in the chaotic regime, the GC results for both observables depart from those obtained 
with the reduced and diagonal density matrices as $R$ increases. This is understandable because the GC results are 
obtained for a system with $L-R$ sites and open boundary conditions, while the other two are obtained for a system 
with $L$ sites from which $R$ sites are either traced out or cut off. Hence, the more sites one cuts off (the larger the
value of $R$), the smaller is the system remaining. Finite size effects are then expected to be stronger, in particular 
as the values of $t',V'$ increase. It is also expected that they should vanish in the thermodynamic limit.

\begin{figure}[!bt]
\includegraphics[width=0.47\textwidth]{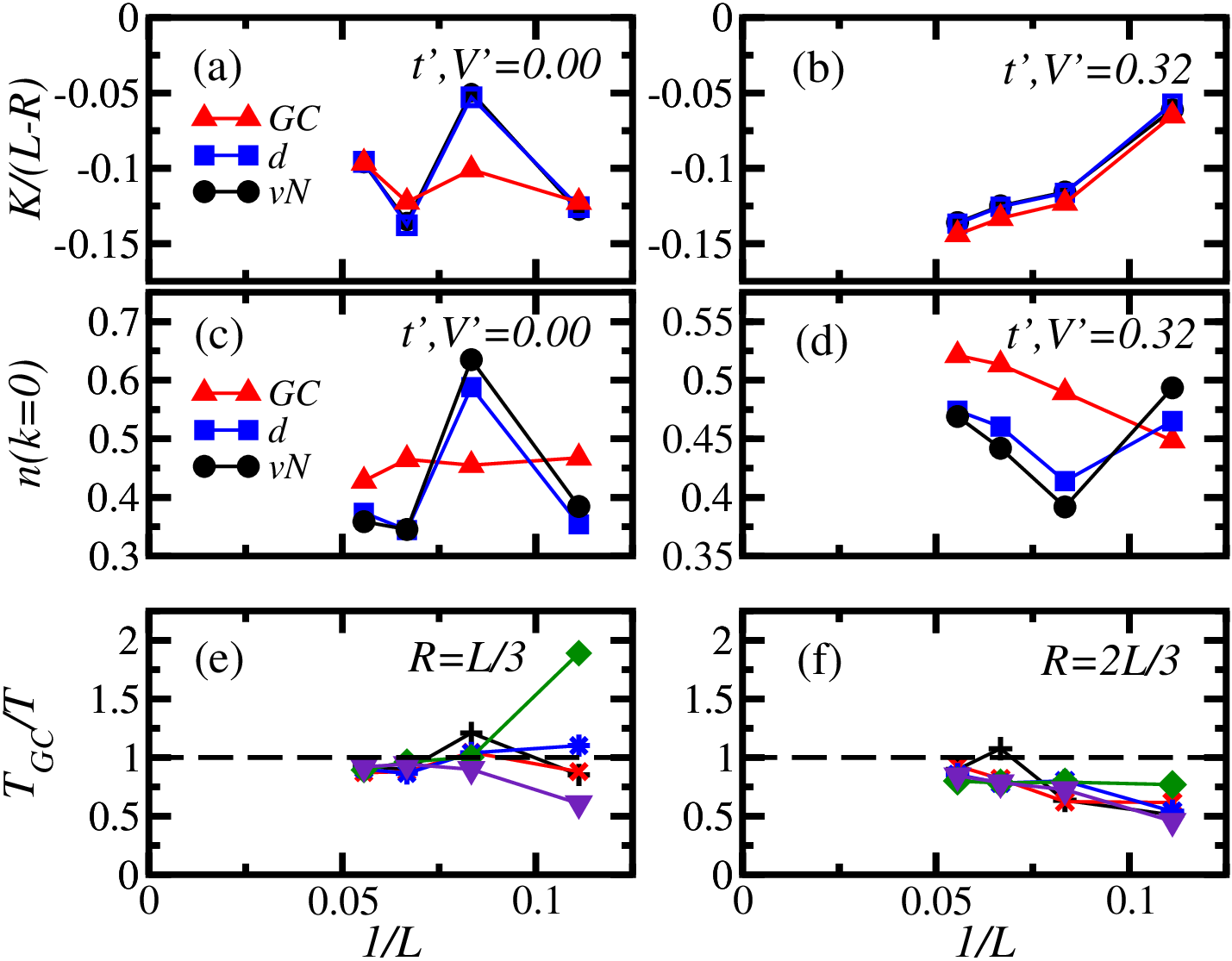}
\vspace{-0.25cm}
\caption{(Color online) (a)--(d) Observables vs $1/L$ for $R=L/3$. 
(e),(f) Ratio between the grand-canonical temperature and the temperature vs $1/L$. 
(Black) Plus: $t'=V'=0.04$; (red) cross: $t'=V'=0.08$; (blue) star: $t'=V'=0.16$; 
(green) diamond: $t'=V'=0.32$; (purple) down-triangle: $t'=V'=0.64$. In all panels: $T=4$.}
\label{fig:OL}
\end{figure}

In Figs.~\ref{fig:OL}(a)--\ref{fig:OL}(d), we show the scaling behavior of the two observables from Fig.~\ref{fig:Ocut} 
for $R=L/3$. It is apparent that, in the chaotic regime [Figs.~\ref{fig:OL}(b) and \ref{fig:OL}(d)], the observables calculated 
in the three ensembles approach each other with increasing system size. (Results for other temperatures and observables in the 
chaotic domain are presented in Ref.~\cite{supp}). In the integrable limit [Figs.~\ref{fig:OL}(a) and \ref{fig:OL}(c)], 
large fluctuations prevent us from reaching general conclusions. This is a regime that needs to be studied by other means, 
which may allow one to do a proper extrapolation to the thermodynamic limit.  

A final important property of the mixed states 
studied here is that their effective thermodynamic temperature ($T_\text{GC}$) is the same as the effective temperature 
$T$ of the many-body eigenstate from which they are obtained. This is shown in Figs.~\ref{fig:OL}(e) and \ref{fig:OL}(f), 
where we plot $T_\text{GC}/T$ vs $1/L$ for $R=L/3$ [\ref{fig:OL}(e)] and $R=2L/3$ [\ref{fig:OL}(f)], and for different 
values of $t',V'$. The two temperatures are close (except for the smallest systems sizes) and further approach each other as $1/L$ 
decreases, which is consistent with the idea of eigenstate thermalization where a single eigenstate exhibits 
thermal behavior \cite{eth,rigol08}.

%%%%%%%%%%%%%%%%%%% CONCLUSIONS %%%%%%%%%%%%%%%%%%%%%%%%%%%%%%%%%%%%%%%%
{\em Conclusions.--} We have found that, away from integrability, the diagonal and grand canonical entropies are 
very close to each other, even when a small portion (probably an infinitesimal fraction) of the lattice is cut off.
They further approach each other with increasing system size. In contrast to $S_d$, $S_\text{vN}$ only approaches 
$S_\text{GC}$ when a large part of the original system is traced out. This behavior reflects the non-diagonal structure of 
the reduced density matrix when only small fractions of the lattice are traced out. Led by these results, we identified two 
different regimes: weak typicality, where the diagonal ensemble develops a thermal structure, and strong (canonical) typicality, 
where the density matrix of the reduced system is thermal. Close to and at integrability further studies are necessary. 
Our results exhibit large fluctuations depending on the state selected and the values of $t',V'$.

We also studied experimentally measurable few-body observables as determined from the reduced, diagonal, and grand-canonical 
density matrices. We find that they all lead to the same results as the system size increases, no matter how many sites are traced out. 
This demonstrates that the extra information contained in the off-diagonal elements of the reduced density matrix is irrelevant 
to those observables. Therefore, it is plausible to conjecture that the expectation values of {\em physical} observables in 
subsystems of any size obtained from generic many-body eigenstates can be entirely obtained from standard statistical ensembles. 
Our results show that even in relatively small systems, both procedures, tracing out the ``bath'' or cutting it off and letting 
the subsystem thermalize, lead to the same outcome.

\begin{acknowledgments}
This work was supported by the NSF under grant DMR-1147430 (L.F.S.), the Office of Naval 
Research  (M.R.), the NSF under grant DMR-0907039, the AFOSR under grant FA9550-10-1-0110, 
and the Sloan Foundation (A.P.). We thank J. Deutsch for useful discussions.
\end{acknowledgments}

{\it Note added.---} After completion of this work, we have learned that the strong (canonical) typicality 
has been recently verified in translationally invariant systems \cite{joshpreprint}.

%%%%%%%%%%%%%%%%%%%%%%%%%%%%%%%%%% SUPPLEMENT %%%%

\onecolumngrid

\vspace*{1.1cm}

\begin{center}

{\large \bf Supplementary material for EPAPS
\\Weak and strong typicality in quantum systems}\\

\vspace{0.6cm}

Lea F. Santos$^1$, Anatoli Polkovnikov$^2$, and Marcos Rigol$^3$\\

$^1${\it Department of Physics, Yeshiva University, New York, NY 10016, USA}

$^2${\it Department of Physics, Boston University, Boston, MA 02215, USA}

$^3${\it Department of Physics, Georgetown University, Washington, DC 20057, USA}

\end{center}

\vspace{0.6cm}

\twocolumngrid

\section{Entropies}

\begin{figure}[!htb]
\includegraphics[width=0.45\textwidth]{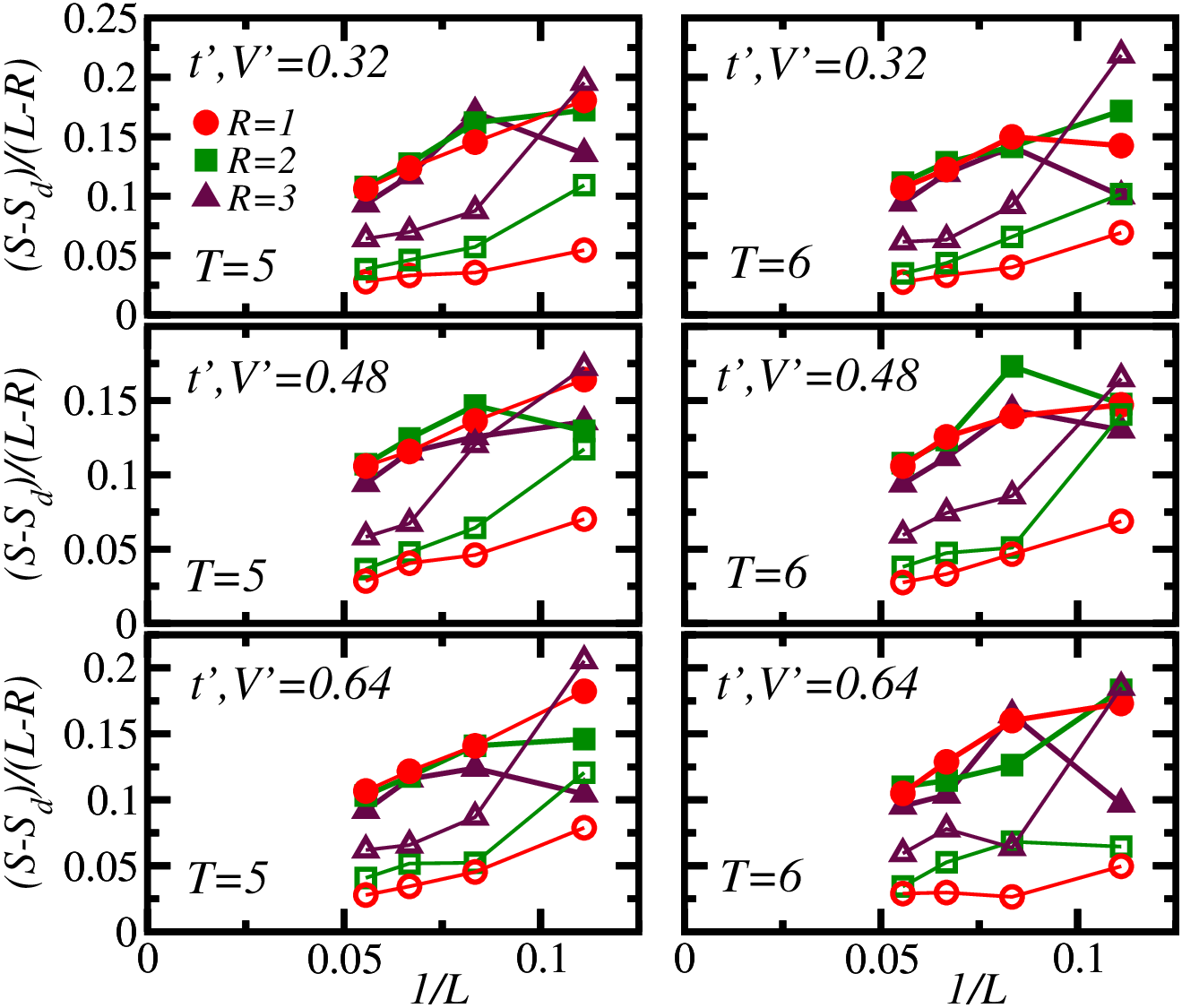}
%\vspace{-0.25cm}
\caption{(Color online) $S_\text{GC} - S_d$ per site vs $1/L$ for the three smallest values of $R$. The values of $t',V'$ are indicated and they imply chaoticity.}
\label{fig:supp01}
\end{figure}

Figure \ref{fig:supp01} provides further support to our statement that $S_d$ approaches $S_\text{GC}$ as the system size increases. This fact holds even for the smallest value $R=1$, suggesting that in the thermodynamic limit, the diagonal density matrix becomes a thermal density matrix when an infinitesimally small part of the system is traced out. 

\section{Observables}

\begin{figure}[htb]
\includegraphics[width=0.47\textwidth]{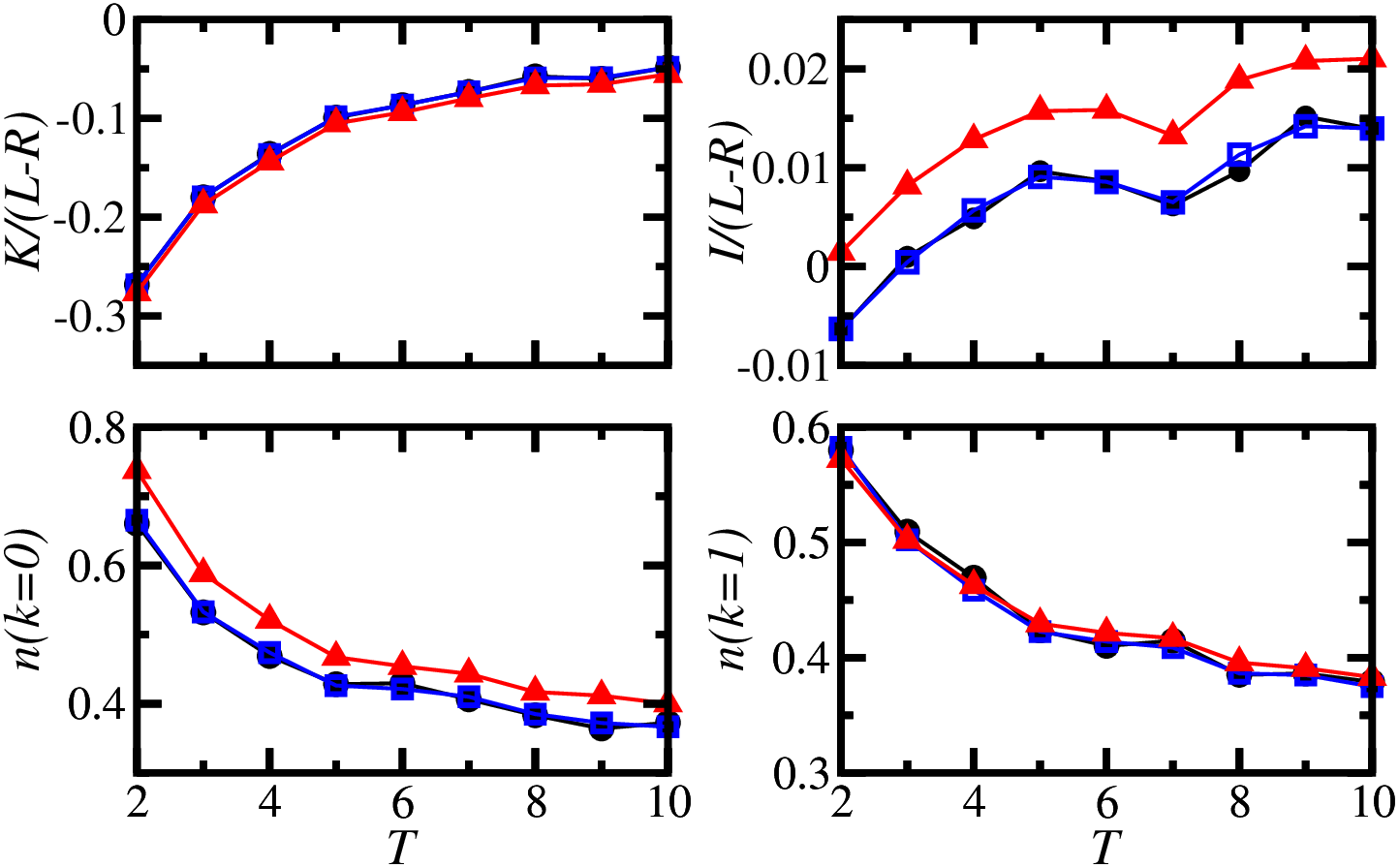}
%\vspace{-0.25cm}
\caption{(Color online) Interaction energy per site, kinetic energy per site, and momentum distribution function for $k=0$ and $k=1$ vs temperature; $L=18$; $R=L/3$; $t'=V'=0.32$. Triangles: $O_\text{GC}$, squares: $O_d$, and circles: $O_\text{vN}$.}
\label{fig:supp02}
\end{figure}

Figure \ref{fig:supp02} reinforces that the expectation values of few-body observables obtained with the reduced, diagonal, and grand canonical density matrices are comparable, independently of the regime, the number of sites traced out, and $T$. In the figure we show the behavior of physical observables vs temperature. We consider the kinetic energy, the momentum distribution function for $k=0$ and $k=1$, and the interaction energy:
\begin{eqnarray}
\hat{I}&=&V\sum_{i=1}^{L-R-1} \left(\hat{n}_i^b -\frac{1}{2} \right)\! \left(\hat{n}_{i+1}^b -\frac{1}{2}\right) 
\label{IE}
\\
&+&V'\sum_{i=1}^{L-R-2} \left(\hat{n}_{i}^b -\frac{1}{2}\right) \! \left(\hat{n}_{i+2}^b -\frac{1}{2}\right).
\nonumber
\end{eqnarray}

The results for $O_d$ and $O_\text{vN}$ are very similar throughout. They indicate that the information contained in the off-diagonal elements of the reduced density matrix is therefore irrelevant to few-body observables. We note that while $O_\text{GC}$ is also very close to $O_d$ and $O_\text{vN}$, the GC results for $n(k=0)$ and $I$ are slightly 
more separated, since these observables are more sensitive to finite size and boundary effects than $K$.
%% How can n(k=1) be explained? GC is very close in this case. 

\onecolumngrid

\begin{figure}[!htb]
\includegraphics[width=0.6\textwidth]{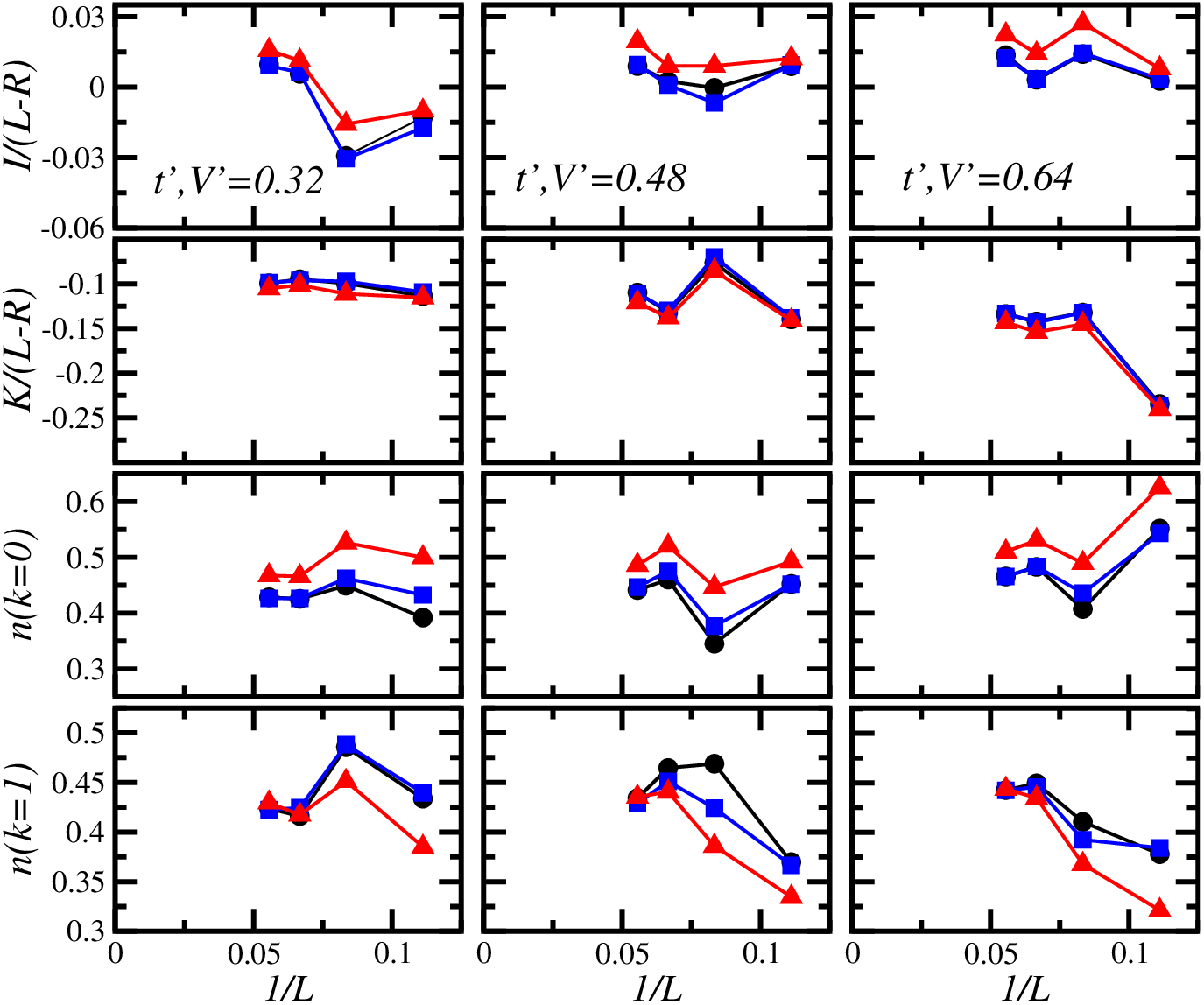}
%\vspace{-0.25cm}
\caption{(Color online) Interaction energy per site, kinetic energy per site, and momentum distribution function for $k=0$ and $k=1$  vs $1/L$; $T=5$.
Triangles: $O_{GC}$, squares: $O_d$, and circles: $O_{vN}$; $R=L/3$. The values of $t',V'$ for each column are indicated at the top panels.}
\label{fig:supp03}
\end{figure}

The scaling behavior of the observables considered here are shown in Fig.~\ref{fig:supp03} for $R=L/3$ and values of $t',V'$ in the chaotic region. In this regime, the observables calculated in the three ensembles approach each other with increasing system size.

\vskip 0.3 cm

\begin{thebibliography}{15}
\bibitem{weights} This conjecture is consistent with the Eigenstate Thermalization Hypothesis \cite{rigol08} implying that in ergodic systems all eigenstates within a small energy window are similar and hence the probabilities to excite them during a quench should be also similar.
\end{thebibliography}
\end{document}